\documentclass{einformatyka}
\usepackage[cp1250]{inputenc} 
\usepackage[T1]{fontenc}
\usepackage{graphicx} 
\usepackage{lscape}
\usepackage{rotating}
\usepackage{tikz}
\usepackage{subfigure}
\usepackage{dblfloatfix}
\usepackage{appendix}
\usepackage{todonotes}
\usepackage{alphalph}
\usepackage{manyfoot}
\usepackage{framed}
\usepackage{lineno}
\usepackage{float}
\usepackage{longtable}
\usepackage{multirow}

\DeclareNewFootnote{B}

\begin{document}

\title{Survey research in software engineering: problems and strategies}

\begin{articleinfo} 

\begin{authorgroup} 
	\begin{author} 
	\firstname{Ahmad Nauman}
	\surname{Ghazi}	
	\orgname{Blekinge Institute of Technology} 
	\orgdiv{Department of Software Engineering} 
	\email{nauman.ghazi@bth.se} 
\end{author}
	
	
\begin{author} 
	\firstname{Kai}
	\surname{Petersen}	
	\orgname{Blekinge Institute of Technology} 
	\orgdiv{Department of Software Engineering} 
	\email{kai.petersen@bth.se} 
\end{author}

\begin{author} 
	\firstname{Sri Sai Vijay Raj}
	\surname{Reddy}	
	\orgname{Blekinge Institute of Technology} 
	\orgdiv{Department of Software Engineering} 
	\email{srre15@student.bth.se} 
\end{author}

\begin{author} 
	\firstname{Harini }
	\surname{Nekkanti}	
	\orgname{Blekinge Institute of Technology} 
	\orgdiv{Department of Software Engineering} 
	\email{hana15@student.bth.se} 
\end{author}

\end{authorgroup}

\keywordset{Empirical software engineering, surveys} 
\begin{abstract}[en] 
\textbf{Background:} The need for empirical investigations in software engineering is 
growing. Many researchers nowadays, conduct and validate their solutions using 
empirical research. Survey is one empirical method which enables 
researchers to collect data from a large population. Main aim of the survey is to 
generalize the findings. \\
\textbf{Aims:} In this study we aim to identify the problems researchers face during 
survey design, and mitigation strategies. \\
\textbf{Method:} A literature review as well as semi-structured interviews 
with nine software engineering
researchers were conducted to elicit their views on problems and mitigation strategies. The researchers are all focused on empirical software engineering. \\
\textbf{Results:} We identified 24 problems and 65 strategies, structured according to the survey research process. The most commonly discussed problem was sampling, in particular the ability to obtain a sufficiently large sample. To improve survey instrument design, evaluation and execution recommendations for question formulation and survey pre-testing were given. The importance of involving multiple researchers in the analysis of survey results was stressed. \\
\textbf{Conclusions:} The elicited problems and strategies may serve researchers during the design of their studies. However, it was observed that some strategies were conflicting. This shows that it is important to conduct a trade-off analysis between strategies. 
\end{abstract}

\end{articleinfo}

\section{Introduction}
\label{sec:intro}

Surveys are a frequently used method in the the software engineering context. Punter et 
al.\cite{punter_conducting_2003} highlighted the increased usage of surveys over 
case-study and experiments.

Surveys are one of the empirical investigation method which is used to collect data from a large population \cite{kasunic_designing_2005}. Surveys have been 
characterized by different authors: Pfleeger highlights that a \textit{``survey is often an 
investigation performed in retrospection \cite{pfleeger_experimental_1995}''}; Babbie adds 
that \textit{``surveys aim is to understand the whole population depending on the sample 
drawn”} \cite{babbie_survey_1973}. Fink \cite{fink_how_2012} states that \textit{``surveys 
are useful for analyzing societal knowledge with individual knowledge''}. Wohlin et al. 
highlight that \textit{``many quantifiable and processable variables can be collected using 
a survey, giving a possibility for constructing variety of explanatory models''} 
\cite{wohlin_experimentation_2012}; Fowler \cite{fowler_jr_survey_2013} states that 
\textit{``statistical evidences can be obtained in a survey''}. and Dawson adds that 
\textit{``surveys draw either qualitative or quantitative data from population''} 
\cite{dawson_projects_2005}.

Stavru \cite{stavru2014critical} critically reviewed surveys and found limitations in relation 
to the definition of the sampling frame, description of the sampling method and the 
definition of the actual sample. Furthermore, the response rate was rarely identified. 
Sampling-related aspects were most highly prioritized as issues\cite{stavru2014critical}. 
Given the limitations in the agile literature there is a need to further explore the use of 
surveys and understanding how they were conducted in the software engineering 
context \cite{stavru2014critical}.  Stavru \cite{stavru2014critical} also points to the need of frameworks to 
evaluate survey research as these were not available in the software engineering literature 
(cf. \cite{stavru2014critical}). Researchers themselves recognize that they are facing 
problems when conducting surveys, highlighting problems such as limited generalizability, 
low response rate, survey reliability, etc.\cite{de_mello_would_2013}, 
\cite{gorschek_large-scale_2010}, \cite{zhang_systematic_2013}, 
\cite{galster_exploring_2014}, \cite{nurdiani_risk_2011}, \cite{yang_survey_2016}. The 
reason for researchers facing problems could be either he/she is unaware of the problems 
or they lack strategies to overcome the problems in the survey process. In both the cases the 
outcome of surveys is unreliable (cf. \cite{pfleeger_principles_2001}). 

Thus, in this study the main focus is on identifying the problems researchers face and 
document in the surveys they are executing and the mitigation strategies they report. In 
particular, the following contributions are made: 
\begin{itemize}
	\itemsep0em
	\item C1: Identify the problems researchers in software engineering face when 
	conducting survey research.
	\item C2: Identify mitigation strategies. 
\end{itemize}

The contributions are achieved through the review of literature combined with an 
interview-study has been conducted with nine subjects. In the literature review we 
focused on existing surveys and elicited problems observed as well as mitigation 
strategies reported in them. A traditional literature review has been used. The interview 
study was based on convenience sampling and face-to-face interviews. Thematic analysis 
has been used to analyze the results of the interviews. 

The remainder of the article is structured as follows: Section \ref{sec:background} presents the background on survey research by explaining the general process of conducting survey research. Section \ref{sec:relatedwork} presents the related work where problems as well as strategies were elicited from existing guidelines as well as primary survey studies conducted in software engineering. Section \ref{sec:method} explains the research design for the interview study conducted. The interview results are thereafter shown in Section \ref{sec:results}. Section \ref{sec:discussion} discusses the findings from the literature study and the interviews. Section \ref{sec:conclusion} concludes the paper.

\section{Background on the survey research method}
\label{sec:background}

\begin{figure}[h]
	\centering
	\includegraphics[height=10cm, width=4cm]{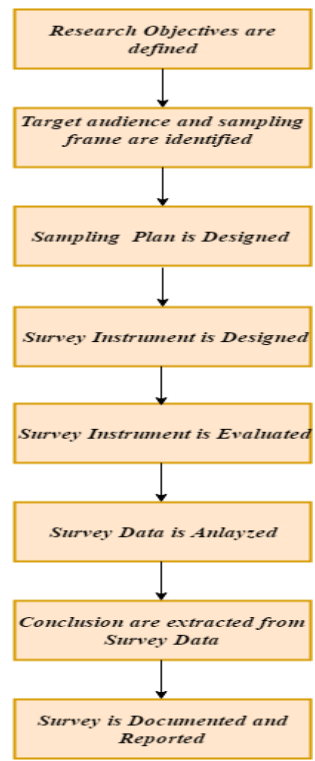}
	\caption{Eight Steps of a Survey}\label{surveysteps}
\end{figure}

Robson and McCartan\cite{robson_real_2016} define the survey methodology as 
\textit{``a fixed design which is first planned and then executed''}.  Molleri et al. reviewed 
the steps of survey research guidelines for software engineering. Commonly defined 
steps are highlighted in Figure \ref{surveysteps}.

\subsection{Research objectives are defined}
The initial step is to identify the research objectives. They help to set the required 
research scope and context for framing the research questions. While identifying the 
research objectives it is essential to throw light on certain issues apart from just 
identifying the research questions. The following reflective questions should be checked 
when defining the research objectives \cite{kasunic_designing_2005}: 
\begin{itemize}
	\itemsep0em
	\item What is the motivation behind Survey?
	\item	What are the resources required to accomplish the survey's goals?
	\item	What are the possible areas which are close to the research objectives that were 
	left uninvestigated? 
	\item What is the targeted respondent population of survey?
	\item	How will the data obtained from survey be used? \cite{kasunic_designing_2005} 
	\cite{kitchenham_personal_2008} \cite{ciolkowski_practical_2003} 
\end{itemize}
While defining the research objectives for a survey, the related work pertaining to that 
particular field must be considered. The knowledge about similar research helps 
researchers to narrow down the objectives. 

Wohlin et al. \cite{wohlin_experimentation_2012} clearly defines the purpose ( objective or 
motive)  for conducting a survey . Based on the objective, any survey falls into one of the 
below three categories:
\begin{itemize}
	\itemsep0em
	\item Descriptive Surveys are conducted with the intention of explaining traits of a 
	given population. For example, they describe which development practices are used in 
	practice. 
	\item	Explanatory Surveys investigate cause-effect relationships. For example, they try 
	to explain why a specific software development practice is not adopted in practice.
	\item	Exploratory Surveys helps the researcher's to look at a particular topic from a 
	different perspective. These surveys are generally done as a pre-study. They help to 
	identify unknown patterns. The knowledge obtained from this pre-study will serve as a 
	foundation to conduct descriptive or exlanatory surveys in the future 
	\cite{wohlin_experimentation_2012}. 
\end{itemize}

\subsection{Target audience and sampling frame are identified}
The identification of the target population implies the establishment of a targeted audience. 
The target audience selection must be driven by the research objectives.The survey 
instrument design must be designed from the respondent's perspective, which requires a 
clear definition of the population and target audience. Similarly, the rule must be applied 
while selecting the method of surveying (questionnaire or interviews) 
\cite{kasunic_designing_2005}.  

The target audience is generally selected from the overall population, if they are attributed 
with distinct values. The sample is selected from the sampling frame comprising of the 
possible respondents from the population. Populations can be categorized into 
sub-populations based on distinguishing attributes, which may be utilized for stratified or 
quota sampling \cite{thompson_sampling_2002}. 
Four basic problems of sampling frames are identified in \cite{kish_survey_1965} which 
are: \textit{``missing elements, foreign elements, duplicate entries and group based 
clusters''}. 

\subsection{Sample plan is designed}
Sampling is the process of selecting a sample for the purpose of studying the 
characteristics of the population. That is, sampling is needed to characterize a large 
population \cite{joseph_jr_hair_2009}. Sampling is mainly divided in two types 
\cite{kitchenham_preliminary_2002}, namely probabilistic and non-probabilistic sampling.

\emph{Probabilistic Sampling:} Each member of the population has a non-zero probability of 
being selected.  Below are the three types of probabilistic sampling techniques 
\cite{thompson1992sampling}:
\begin{itemize}
	\itemsep0em
	\item 	Random Sampling: Members of the sampling frame are selected at random.
	\item	Systematic Sampling: A sampling interval is determined ($k$) and every $kth$ 
	element is chosen from the sampling frame.
	\item Stratified Sampling: The sampling frame is divided into different groups (e.g. 
	based on experience level of developers in an experiment) and the subjects are chosen 
	randomly from these groups. 
\end{itemize}

\emph{Non-probabilistic Sampling:} Member selection in this case is done in some 
non-random order. Below are the types of non-random sampling techniques 
\cite{fowler_jr_survey_2013}, \cite{kasunic_designing_2005}:
\begin{itemize}
	\itemsep0em
	\item 	Convenience Sampling: Subjects are selected based on accessibility. Examples 
	are the utilization of existing contact networks or accessing interest groups (e.g. 
	LinkedIn) where subjects are available that are clearly interested in the subject of the 
	survey.
	\item	Judgment Sampling: The sample is selected through the guidance of an expert. 
	For example, a company representative for a company-wide survey may choose the 
	subject best suited to answer the survey due to their expertise. 
	\item	Quota Sampling: Similar to stratified sampling the sample is divided into gropus 
	with shared traits and characteristics. However, the selection of the elements is not 
	conducted in a random manner.
	\item	Snowball Sampling: Existing subjects of the sampling frame are utilized to recruit 
	further subjects. 
\end{itemize}

\subsection{Survey instrument is designed}
Survey outcomes directly depend on how rigorous the survey has been designed. 
Questions (such as open and closed questions) are designed, and different question 
types are available (e.g. Likert-scale based questions). The factors which needs to be 
considered while designing surveys have been discussed by Kasunic 
\cite{kasunic_designing_2005}.

\subsection{Survey Instrument is Evaluated}
After the Survey Instrument has been designed, it needs to be evaluated to find out if 
there are any flaws. To determine a questionnaire’s validity a preliminary evaluation is 
conducted. Examples of different evaluation methods are:
\begin{itemize}
	\itemsep0em
	\item Expert Reviews \cite{stahura_stanley_2005}.
	\item Focus Groups \cite{stahura_stanley_2005}. 
	\item Cognitive Interviews 
	\cite{stahura_stanley_2005}\cite{haeger_using_2012}\cite{litwin_how_1995}.
	\item Experiment \cite{moore_using_2004}.
\end{itemize}

\subsection{Survey data is analyzed}
The obtained survey data is analyzed in this step. The data analysis depends on the type 
of questions used in the survey. 
\begin{itemize}
	\itemsep0em
	\item Common methods to analyze the results of open-ended questions are 
	phenomenology, discourse analysis, grounded theory, content analysis and thematic 
	analysis \cite{elo_qualitative_2008}, \cite{hsieh_three_2005}, 
	\cite{boustedt_methodology_2008}, \cite{ghanam_making_2012}, 
	\cite{sharp_tensions_2004}. 
	
	\item For closed-ended questions, quantitative analysis can be employed. Methods such 
	as statistical analysis, hypothesis testing, and data visualizations can be employed to 
	analyze the closed-ended questions \cite{wohlin_experimentation_2012}.
\end{itemize}

With regard to the analysis process Kitchenhamm and Pfleeger 
\cite{kitchenham_principles_2002} suggest the following activities:
\begin{enumerate}
	\item \emph{Data Validation:} Before evaluating the survey results, researchers must 
	first check the consistency and completeness of responses. Responses to ambiguous 
	questions must be identified and handled. 
	\item	\emph{Partitioning of Responses:} Researchers need to partition their responses 
	into subgroups before data analysis. Partitioning is generally done using the data 
	obtained from the demographic questions. 
	\item	\emph{Data Coding:} When statistical packages cannot handle the character 
	string categories of responses, researchers must convert the nominal and ordinal scale 
	data.
\end{enumerate}
Wohlin et al.\cite{wohlin_experimentation_2012} describes the first step of quantitative 
interpretation where data is represented using descriptive statistics visualizing the central 
tendency, dispersion, etc. The next step is data set reduction where invalid data points 
are identified and excluded. Hypothesis testing is the third step.

\subsection{Conclusions extracted from survey data}
After the outcomes have been analyzed, conclusions need to be extracted from them. A 
critical review and an evaluation must be done on the obtained outcomes. Thus validity, 
reliability and risk management should be evaluated when presenting conclusions. Every 
research has threats, but the main motive is to identify them at the early stages and try to 
reduce them. Threats may be completely mitigated by research design decisions, while 
other threats remain open or may only be partially reduced. To handle such threats, it is 
advised that more than one method must be used to achieve a research objective for 
reducing the impact of a particular threat \cite{ciolkowski_practical_2003} 
\cite{robson_real_2016}. 

\subsection{Survey documented and reported}
\label{sec:surveyreporting}

The documentation of the survey design is updated iteratively as the research process 
progresses. Different elements of documentation include RQ's, objectives, activity 
planning, sample method design, data collection, data analysis methods, etc. This 
documentation is referred to as \textit{``questionnaire specification''} by 
\cite{kitchenham_personal_2008}, while it is named a \textit{``survey plan''} by Kasnunic 
\cite{kasunic_designing_2005}. 

The last step is the reporting of the analysis and conclusion. Even though the survey 
methodology is administered sequentially, results reporting might vary depending on the 
targeted readers (e.g. researchers or practitioners). Since the interests of audiences 
differ, Kasunic \cite{kasunic_designing_2005} recommend conducting an audience 
analysis. Stavru \cite{stavru2014critical} evaluated existing surveys in software 
engineering and identified the most critical elements to be reported in surveys. The most 
critical elements were: 
\begin{itemize}
	\itemsep0em
	\item The sampling frame and the number of elements in the sampling frame. 
	\item The strategy of sampling from the sampling frame
	\item The size of the sample
	\item The target population
	\item The response rate
	\item Assessment of the trustworthiness of the survey
	\item Execution of the survey (research steps)
	\item Concepts and theories used (e.g. variables studied)
	\item The design of the survey
\end{itemize}

\section{Related Work}
\label{sec:relatedwork}

\subsection{Guidelines for survey research in software engineering}

Molleri et al. \cite{MolleriPM16} surveyed the literature to identify guidelines for survey 
research. Three literature sources 
\cite{kasunic_designing_2005,linaker2015guidelines,kitchenham2002principles} presented 
the overall survey process, while several studies focused on individual parts of the 
process (e.g. only planning and execution). Overall, Molleri et al.\cite{MolleriPM16} found 
that the different processes comprise of similar steps, while they have different 
granularities. 

The article by Kasunic \cite{kasunic_designing_2005} described guidelines for 
conducting a survey. The author describes each step in the survey process and formed 
the basis to structure the background reported in this paper (Section 
\ref{sec:background}). 

In addition to overall processes prescribed for survey research several guidelines focused 
on specific aspects of survey research.

Punter et al.\cite{punter_conducting_2003} presented guidelines focusing mainly on 
online-surveys. They have drafted a set of guidelines to perform online survey from their own 
experiences of conducting five on-line surveys. They highlighted that data obtained from 
online surveys is easy to analyze as it is obtained in the expected format while 
paper-based forms are error prone. Oline surveys track the responses of invited 
respondents and log the details of those who actually answered the survey, which allows 
to more easily follow up and increase response rates. Punter et 
al.\cite{punter_conducting_2003} argued that online surveys help to gather more 
responses and ease the disclosure of the results obtained.

Low response rates are a common problem for any survey, which was identified by Smith 
et al. \cite{smith_improving_2013}. Based on their expertise and the 
existing literature, they performed a post-hoc analysis on previously conducted surveys 
and came up with factors to improve participation rate. They even specified the limitations 
of the obtained results stating that  \textit{``an increase in participation doesn't mean the 
results become generalizable''} \cite{smith_improving_2013}.

Pertaining to the survey sampling, Travassos et al. \cite{de_mello_investigating_2015} propose a framework consisting of target population, sampling frame, unit 
of observation, unit of attribute and an instrument for measurement. Ji et al. 
\cite{ji_lessons_2008} have conducted surveys in China and addressed the issues 
relating to sampling, contacts with respondents and data collection, and validation issues. 
Conradi et al. \cite{conradi_reflections_2005} have highlighted the problem of method 
biases, expensive contact processes, problems with census type data, and national variations by 
performing an industrial survey in three countries – Norway, Italy and Germany. This is the 
first study in software engineering which used census type data. The problem of 
replications of surveys was highlighted by Rout et al.\cite{cater-steel_addressing_2005} 
who replicated a European survey, which was administered in Australian software 
development organizations.

\subsection{Problems and strategies}

The problems and strategies in literature are structured according to the steps presented 
in Figure \ref{surveysteps}. We first present the problems (LP**) and the strategies (LS**) mentioned in 
the literature that were directly linked to the problems by the authors.

\subsubsection{Target audience and sampling frame definition and sampling plan}

\textbf{LP01: Insufficient Sample Size:} Insufficient sample size is the major threat for any 
software engineering survey. Meaningful statistical evidences cannot be obtained even 
when the parametric tests are applied on to a particular sample due to insufficient size 
\cite{martini_multiple_2016}, \cite{neto_25_2013}. One of the main aims of Surveys is to 
generalize findings to a larger population. Generalizability increases survey’s confidence. 
Small sample size is attributed as the main cause for the lack of generalizability. If 
generalizability is not possible then the whole aim of the survey is not achieved 
\cite{de_mello_would_2013} \cite{yamashita_surveying_2013} 
\cite{gorschek_large-scale_2010} \cite{zhang_systematic_2013} 
\cite{cavalcanti_towards_2013} \cite{garousi_survey_2015}. As Kitchenham and Fleeger 
\cite{kitchenham_principles_2002} describe, inadequate sample size negatively impacts the 
survey outcomes in two ways. Firstly, deficient sample size leads to results that do not 
show any statistical significance. Secondly poor sampling of clusters reduces the 
researcher’s ability to compare and contrast various subsets of the population.  

Reasons are small sample sizes are busy schedules of the respondents 
\cite{galster_exploring_2014}\cite{akbar_framework_2012}, poorly designed survey layout, 
lack of awareness about survey and long surveys \cite{galster_exploring_2014}. Conradi et 
al. \cite{ji_lessons_2008} explained the impact of culture on response rates. They argued 
that socio-economic positions of the respondents might hinder their willingness to 
answers.  Authors showed that collectivism had direct influence on the information 
sharing, where people are not interested in sharing information outside their group (i.e. 
with researchers). Several solutions have been proposed in the literature: 
\begin{itemize}
	\itemsep0em
	\item \textit{LS01: Use personal contact network:} The personal contact network is 
	used to recruit respondents \cite{galster_exploring_2014}, \cite{vukelja_are_2007}, 
	\cite{nurdiani_risk_2011}, \cite{de_mello_would_2013}, \cite{akbar_framework_2012}. 
	
	\item \textit{LS02: Cultural awareness: }This issue can be handled by carefully designing 
	questionnaire being aware of the cultures of the respondents \cite{ji_lessons_2008}.
	
	\item \textit{LS03: Use probabilistic sampling:} If researchers aim is to generalize to a 
	target population, then probabilistic sampling must be considered 
	\cite{kitchenham_principles_2002}.
	
	\item \textit{LS04: Use of convenience sampling:} Garousi et 
	al.\cite{garousi_cross-factor_2016} describe the motivation for researchers selecting 
	convenience sampling over other techniques, highlighting that convenience sampling is 
	less expensive and troublesome. 
	
	\item \textit{LS05: Evaluate the trustworthyness of the sample 
	\cite{stavru2014critical}:}  Different ways for calculating the sample size depending on 
	the size of the population have been proposed \cite{kasunic_designing_2005}. 
	
	\item \textit{LS06: Reciprocity:} Researchers can induce reciprocity (respondents 
	answer more than once, e.g. for different projects) by giving rewards. Smith et al 
	\cite{smith_improving_2013} were not sure whether this practice was actually useful in 
	software engineering domain as it may introduce a bias in the results.
	
	\item \textit{LS07: Consistency:} It is the nature of humans to experience cognitive 
	pressure when they are not performing the promised deeds.  This characteristic can 
	induce more responses for a survey \cite{smith_improving_2013}.
	
	\item \textit{LS08: Authority and Credibility:} The compliance for any kind of survey can 
	be increased by the  credibility of the person who is administering the survey. 
	Researchers can utilize this benefit by providing the official designations like Professor 
	or PhD in the signature of the survey request mail \cite{smith_improving_2013}. 
	
	\item	\textit{LS09: Liking:} Respondents tend to answer the surveys from known 
	persons. The responsibility of gaining trust lies with the researchers 
	\cite{smith_improving_2013}.
	
	\item	\textit{LS10: Scarcity:} It is the human nature to react fast when something is 
	scarce, research can increase the survey’s response rate by convincing about the 
	survey's uniqueness. \cite{smith_improving_2013}.
	
	\item	\textit{LS11: Brevity:} Respondents tend to answer shorter surveys compared to 
	lengthy ones. Researcher should address the number of questions at the start of 
	survey, a progress bar must be placed to help respondents know the survey progress. 
	Usage of close ended questions also helps to attract more respondents 
	\cite{smith_improving_2013}.
	
	\item \textit{LS12: Social Benefit:} Authors describe that more respondents finish the 
	survey if it benefits to a large group instead of a particular community. Researchers 
	must convince the respondents that their survey benefits larger population 
	\cite{smith_improving_2013}. 
	
	\item \textit{LS13: Timing:} The time at which an email survey is sent also affects its 
	response rate. A study shows that respondents tend to answer emails right after their 
	lunch \cite{smith_improving_2013}.
	
	\item \textit{LS14: Define clear criteria for sample selection:} Selecting the respondents 
	based on a set of criteria (that are defined at the survey instrumentation stage) can 
	reduce the chances of improper selection \cite{spinola_towards_2012}.
	
	\item \textit{LS15: Third party advertising:} Third party advertising can lead to more 
	survey responses, Bacchelli et al\cite{gousios_work_2016} obtained a 25\% increase in 
	responses rate by following this process. Deuresen at al. \cite{gousios_work_2015} 
	have used customized reports along with third party advertising to increase their 
	response rate.
	
	\item \textit{LS16: Use snowball sampling:} Respondents of the survey are asked to 
	answer and forward it to their colleagues 
	\cite{garousi_cross-factor_2016}\cite{garousi_survey_2015}. 
	
	\item \emph{LS17: Recruit respondents from GitHub:} Testers and coders can be 
	recruited for a survey using GitHub \cite{cito_making_2015}\cite{gousios_work_2015}.
	
	\item \emph{LS18: Provide rewards:} Researchers can attract the respondents by giving 
	rewards like Amazon points or vouchers gifts. They have to be careful about the 
	responses obtained, since respondents might just answer survey for sake of rewards 
	or answer it twice \cite{cito_making_2015}\cite{daka_survey_2014}.
\end{itemize}

\textbf{LP02: Confidentiality Issues:} In some case software engineering researchers 
would like to observe on-going trends in the industry or study about specific industrial 
issues. Though, the software companies do not allow the respondents to take the survey 
due to the issue of confidentiality. This is problem was faced by one of researchers in 
their survey \textit{``their companies would not allow employees to take this survey due 
to concerns about confidentiality''} \cite{ji_lessons_2008}. 
\begin{itemize}
	\itemsep0em
	\item \textit{LS19: Personalized e-mails:} This threat could be mitigated by sending 
	personal emails rather than system generated emails and by having a follow-up with all 
	those respondents till the survey ends \cite{ji_lessons_2008}. Even if this does not 
	handle the issue then it is better to have personal meeting to discuss about the survey. 
\end{itemize}

\textbf{LP03: Gate Keeper Reliability:} A gate keeper (person having all the details of 
employees) from a particular company is contacted by the researcher. The questionnaire 
is then sent to the gatekeeper, then he/she forwards it to respondents in that company. 
Sometimse respondents do not receive questionnaire resulting in the a lower participation 
rate for a survey. 
\begin{itemize}
	\itemsep0em
	\item \textit{LS20: Use IT responsibles for reliable distribution of invitations:} This issue 
	was reported by Conradi et al. in their research. Authors mitigated this problem by 
	contacting IT-Responsible for that particular company for getting respondent details 
	\cite{ji_lessons_2008}.
\end{itemize}

\textbf{LP04: No Practical Usefulness:} Any surveys that does not prove to be useful to 
the respondents, chances are much likely to skip the survey. Authors of 
\cite{torchiano_six_2013} clearly show this in the following lines \textit{``by far the study 
is interesting but to whom are the results useful for?''}. 
\begin{itemize}
	\itemsep0em
	\item \textit{LS21: Explicitly motivate the practical benefit of the survey:} This issue can 
	be handled by motivating the respondents by giving description about survey 
	outcomes and need for answering survey. 
\end{itemize}

\subsubsection{Survey instrument design,  evaluation, and execution}

\textbf{LP05: Flaws in the wording of questions:} Sometimes questions are ambiguous, confusing or leading 
\cite{gousios_work_2016,ernst_measure_2015}.  When survey questionnaire is not clearly 
understood the respondents arrive at wrong conclusions about questions, as a result 
they answer incorrectly \cite{torchiano_six_2013}. Respondents may give two contrary 
answers for the same question, i.e. being inconsistent within the same survey 
\cite{yang_survey_2016}.  This problem can be handled by posing same question in 
different ways \cite{yang_survey_2016}. 
\begin{itemize}
	\itemsep0em
	\item \textit{LS22: Survey pre-test:} Researchers 
	\cite{gousios_work_2016,ernst_measure_2015} pretested the survey with subjects 
	(internally as well as externally with real subjects). 
	\item \textit{LS23: Expert discussions:} Discussions with colleagues and domain experts 
	were also the part of pre-test process. Gorschek et al.\cite{gorschek_large-scale_2010} 
	have also done redundancy check in addition pre-tests and expert discussion to handle 
	the Survey Instrumentation Problems. Authors Travassos et 
	al.\cite{spinola_towards_2012} used external researchers that are not involved in the 
	research and reformulated the questionnaire based on their reviews.
	\item \textit{LS24: Ask the same question in different ways:} Lack of consistency and 
	understanding can be handled by posing same question in different ways 
	\cite{yang_survey_2016}
\end{itemize}

\textbf{LP06: Translation Issues:} Translation issue is one of the common problems faced 
in globally conducted surveys. Avgerio et al. \cite{yang_survey_2016} conducted a global 
survey in Europe and China. The authors posted a questionnaire after translation. As a 
result of a poor translation data loss occurred. It led to misinterpretation by the 
respondents leading to false answers. 

\begin{itemize}
	\itemsep0em
	\item \textit{LS25: Collaboration with international researchers:} This problem can be 
	handled when researchers working same domain of the same origin are involved in 
	translation process. Language issue like accent and sentence formulation can be 
	handled in the same manner\cite{gousios_work_2016}, \cite{ji_lessons_2008}. Solutions 
	are:
\end{itemize}

\textbf{LP07: Biases due to Question-Order Effect:} Question-order effect 
\cite{gousios_work_2016} means that the order of the questions is a confounding factors 
influencing the answers by the subjects. 
\begin{itemize}
	\itemsep0em
	\item \textit{LS26: Order randomization:} This issue can be mitigated by the authors by 
	randomizing the questions of the questionnaire \cite{gousios_work_2016}.
	
	\item \textit{LS27: Natural actions-sequence:} Designed the questionnaire based on a 
	natural actions-sequence helping the respondents in recalling and understanding the 
	questionnaire properly \cite{gousios_work_2015}. 
\end{itemize}

\textbf{LP08: Likert Scale Problems:} A Likert scale is one dimensional in nature, 
researchers mostly use this in surveys with an assumption that respondent's opinions can 
be mapped well to a construct represented by the Likert scale (e.g. team motivation can 
be surveyed, but is a very complex construct). In a realistic scenario this is not true. Some 
respondents might get confused on what responses to pick, settling for the middle 
option. Analyzing the results obtained by higher order Likert scales for analysis posing a 
threat of misinterpretation or data losses \cite{ernst_measure_2015}. 
\begin{itemize}
	\item \emph{LS28: Avoid two-point scales} Researchers should avoid two point Likert 
	scales ‘yes/no’, instead they are advised to use other multi-point scales 
	\cite{cater-steel_addressing_2005}.
\end{itemize}

\textbf{LP09: People Perceptions}: Perception of people answering the survey adversely 
impacts the survey outcome. In software engineering a survey is done to collect the 
attitudes, facts, and behaviors of the respondents.  This issue cannot be mitigated or 
controlled completely \cite{torchiano_six_2013}.

\textbf{LP10: Lack of Domain Knowledge:} A posted survey could be answered by the 
respondents without proper domain knowledge. This leads to misinterpretation of the 
questionnaire resulting in wrong answers  \cite{yang_survey_2016}, 
\cite{cater-steel_addressing_2005}\cite{martini_multiple_2016}, \cite{ji_lessons_2008}. Ji et 
al.\cite{ji_lessons_2008} commented that \textit{``busy executives likely ignore the 
questionnaires, sometimes their secretaries finish the survey. In some case the responses 
obtained are filled with out by the respondents without domain knowledge''}. One solution 
proposed was:
\begin{itemize}
	\item \textit{LS29: Explicitly consider background knowledge in the survey:} Gorschek 
	et al. \cite{gorschek_large-scale_2010} stressed the need for considering the impact of 
	background influence of the subjects on survey results while surveying.  
\end{itemize}

\textbf{LP11: High drop-out rates:} Sometimes respondents start 
answering the surveys, but they lose interest after some time as the survey progresses; 
boredom leads to the low response rate. Lengthy surveys might a reason for the 
respondents to feel bored \cite{galster_exploring_2014}. One obvious solution is:
\begin{itemize}
	\item \textit{LS11: Brevity:} Researcher should limit the number of questions.
\end{itemize}

\textbf{LP12: Time constraints of running the survey:} Time limitations put on surveys as a 
constraint limit the response rate. Smite et al.\cite{nurdiani_risk_2011} showed that time 
limitation is the main factor for respondents not answering questionnaire or taking phone 
interviews. It can be clearly seen from these lines \textit{``all the 13 respondents were 
asked to take part, due to time limitation we obtained only 9 responses.''} Sometimes 
researchers neglect the responses obtained from the actual subjects due to time 
limitation, following lines discuss about this issue \textit{``due to rather low response 
rate and time limits, we have stopped on 33 responses, which covers 13.58\% of the Turin 
ICT sector''} \cite{egorova_evaluating_2009}.

\textbf{LP13: Evaluation Apprehension: } People are not always comfortable being 
evaluated, which affects the outcome of any conducted study 
\cite{wohlin_experimentation_2012}. It is the same case with survey studies, sometimes 
respondents might not be in a position to answers all the questions, instead they shelter 
themselves by just selecting safer options. This affects the survey outcomes. The 
following solution has been proposed:
\begin{itemize}
	\item \textit{LS30: Guarantee anonymity:} Anonymity of subjects reduced this problem 
	of evaluation apprehension  \cite{gorschek_large-scale_2010}.
\end{itemize}

\textbf{LP14: Common biases of respondents:} Bias or one-sidedness is a common 
problem during the survey process. Common types of biases are: 

\underline{Mono-operation Bias:} Sometimes the instrument in survey process might 
under present 
the theory involved, this is called mono-operation bias 
\cite{wohlin_experimentation_2012}. Solutions are:
\begin{itemize}
	\item  \emph{LS24: Ask the same question in different ways:} Framing different 
	questions to address the same topic \cite{gorschek_large-scale_2010}, 
	\cite{martini_multiple_2016}
	\item \emph{LS31: Source triangulation:} Collecting data from multiple sources 
	\cite{gorschek_large-scale_2010}, \cite{martini_multiple_2016}
\end{itemize}

\underline{Over-estimation Bias:} Sometimes the respondents of the survey over-estimate 
themselves, introducing bias into survey results. Mello and Travassos 
\cite{de_mello_would_2013} identified that \textit{``LinkedIn members tend to 
overestimate their skills biasing the results''}. 

\underline{Social Desirability Bias:} There are situations where respondents tend to appear 
in the 
positive light. This might be due the fear of being assessed by the superior authorities. 
This has a lot of influence on survey outcomes. The following strategy is proposed:
\begin{itemize}
	\item \emph{LS30: Guarantee anonymity:} Maintaining the anonymity in responses and 
	sharing the 
	overall survey result after reporting \cite{gousios_work_2016}.
\end{itemize}

\textbf{LP15: Hypothesis Guessing:} This is a construct validity threat where respondents 
guess the expected survey outcomes, they try to base that anticipation (hypothesis) 
towards answering questions either in a positive way or a negative way 
\cite{wohlin_experimentation_2012}. 
\begin{itemize}
	\item \emph{LS32: Stress importance of honesty:} Gorscheck et.al 
	\cite{gorschek_large-scale_2010} tried to mitigate by stressing the importance of 
	honesty in the introduction of the survey by means of a video and a web page.  
\end{itemize}


\textbf{LP16: Respondent Interaction:} This is a conclusion validity threat. During the 
survey process the respondents might interact and thus influence each other.  In small 
surveys this threat has a large impact on the survey outcome, but in case of surveys done 
at large scale the impact gradually decreases \cite{gorschek_large-scale_2010}.

\subsubsection{Data analysis and conclusions}

\textbf{LP17: Eliminating invalid responses:} In large scale surveys, during analysis this 
problem poses a lot of work to the researcher as they need to eliminate all the incorrect 
responses. A strategy is voluntary participation.
\begin{itemize}
	\item \emph{LS27: Voluntary participation:} This problem can be reduced by making the 
	survey strictly voluntary and only collecting data from the respondents who are willing 
	to contribute \cite{yang_survey_2016}.
\end{itemize}

\textbf{LP18: Response Duplication:} A major problem is faced in open-web surveys is 
response duplication, where the same respondent answers the questionnaire more than 
one time \cite{lucassen_use_2016}\cite{ernst_measure_2015}\cite{gousios_work_2015}. 

\textbf{LP19: Inaccuracy in data extraction and analysis}: Inaccuracy in the data extraction 
and analysis might arise when data extraction from the questionnaire and result reporting 
are done by an individual person \cite{ernst_measure_2015}. 
\begin{itemize}
	\item \emph{LS28: Multiple researchers conduct analysis:} Multiple researchers should 
	be utilized when extracting and analyzing the data \cite{ernst_measure_2015}.
	\item \emph{LS29: Check the consistency of coding between researchers:}  Two 
	researchers may check their inter-rater reliability through an analysis using the Kappa 
	statistic \cite{ernst_measure_2015}.
	
\end{itemize}

\subsubsection{Reporting}

\textbf{LP20: Lack of Motivation for sample selection:} Many researchers fail to report 
their motivation for sample selection \cite{stavru2014critical}. 

\textbf{LP21: Credibility:} For the survey methodology to be accepted as credible and 
trustworthy, the research method and results need to be clearly presented 
\cite{stavru2014critical}.

\section{Research Method}
\label{sec:method}

\subsection{Research questions}

We formulated a corresponding research question for each contribution. 
\begin{itemize}
	\itemsep0em
	\item \textit{RQ1:} Which problems do researchers in software engineering report when 
	conducting surveys?
	\item \textit{RQ2:} Which strategies do they suggest to overcome the problems?
\end{itemize}

\subsection{Selection of subjects}

Initially a list of 20 software engineering researchers were chosen to be interviewed. We 
focused on people conducting empirical software engineering research and included early career researchers as well as senior researchers (PostDocs and professors). Request mails 
were sent stating the research purpose and the need for their appointment. We received 
nine positive replies stating their willingness for an interview. The interviews were 
conducted face-to-face. All the interviews were conducted for a time-span of 50 to 90 
minutes. The subjects included four professors, two PostDoc
researchers and three PhD students, as shown in Table \ref{intervieweedetails}. Overall, the table shows that the researchers have substantial experience. 

\begin{table}[!t]
	\centering
	\caption{Interviewee's Details}
	\scalebox{0.7}{
	\begin{tabular}[h]{c c c c c }
		\hline
		\textbf{ID} & \textbf{Position} & \textbf{Research experience (years)} &   \textbf{\#Publications (DBLP)} &  \textbf{Time taken (minutes)} \\ 
		\hline
		1         & Professor            & 32	&170   & 80                  \\ 
		2         & Professor            & 16   & 73& 90                  \\ 
		3         & Professor            & 12   & 70 &60               \\ 
		4         & Professor            & 15   & 37 &40                   \\ 
		5         & Post Doctoral Researcher & 8    & 11       & 60                  \\ 
		6         & Post Doctoral Researcher  & 9   &    18    & 60                 \\ 
		7         & PhD student       & 4     & 4  & 90                 \\ 
		8         & PhD student       & 5    &  10  & 50                 \\ 
		9         & PhD student      & 5     &  17  & 90                   \\ 
		\hline
	\end{tabular}}
	\label{intervieweedetails}
\end{table}

\subsection{Data collection}

Generally, interviews are conducted either way individually or with group of people, focus 
groups \cite{rowley_conducting_2012}. In this research we have conducted individual 
interviews where interviews are done one person at a time. The characteristics of the 
interview that we have conducted are as follows \cite{kajornboon_using_2005}:
\begin{itemize}
	\itemsep0em
	\item \textit{Use of open-ended questions:} Through these questions we aimed for an 
	extended discussion of the topic. In this way interviewees had the freedom of 
	expressing their opinions based on their experiences.
	\item \textit{Semi-Structured format:} We focused on getting an in-depth knowledge of 
	topic thorough interviews. This can be achieved if the interviewer has a set of 
	questions and issues that were to be covered in the interview and also ask additional 
	questions whenever required. Due to this flexibility have chosen semi-structured 
	interviews.
	\item \textit{Recording of responses:} The interviews were audio recorded with 
	interviewees consent . Field notes were maintained by the interviewer which were 
	helpful in the deeper meaning and better understanding of the results.
\end{itemize}

The aim of this interview questionnaire is to investigate the problems faced by
the researchers while conducting surveys in software engineering. This questionnaire
is divided into two sets of questions The first set of questions mainly focuses
on problems that are commonly faced by the researchers like cultural issues, instrument
flaws, validity threats and generalizability issue. The interviewee is
expected to answer these questions from a researcher's perspective. The second
set of questions mainly focuses on problems that a respondent faces
while answering a survey. It also includes the questions asking
for suggestions and recommendations regarding the questionnaire design. The interviewee
(software engineering researcher) is expected to answer these questions
from a respondent's Perspective.

Finally, the questionnaire ends by asking researchers for their strategies to address the 
problems raised earlier. 

The complete questionnaire is can be found in Appendix \ref{a:Interviewguide}. 

\subsection{Data analysis}

We have chosen thematic analysis process to analyze the results obtained during the 
interviews. Although there are many other procedures that can be followed to analyze we 
have a strong reason for opting thematic analysis. The information which needs to be 
analyzed is the information obtained after conducting several interviews.   Since, we were 
analyzing the results obtained from several interviews, we believed that thematic analysis 
will assist in analyzing the information very effectively. In the following part of this 
section, we are going to describe several steps performed during 
analysis\cite{cruzes_recommended_2011}. 

\emph{Extraction of Information:} In this stage, we collect all the data from the transcripts 
prepared from all interviews. As explained above, our transcripts were prepared 
immediately after the interviews. We have made field notes during each and every 
interview to make sure that all the interviewees exact view point and their suggestions 
about our research were penned down during the interview itself. We have collected all 
these information and documented as a part of this data extraction process. We have 
gone through all the interview transcripts several times in order familiarize ourselves 
about the information which we have extracted from our interviews both verbally and 
non-verbally. We made sure that we have a clear idea of all the information which we had 
extracted\cite{cruzes_recommended_2011}.
\emph{Coding of Data:} As a process of coding our data, we have exclusive codes for all 
the interviews we conducted. We started with Interview1, Interview2 and so on. This will 
ensure that our information is segregated according to the interviews which will assist us 
during the later phases of analysis. We also provided coding few concepts which are 
similar for all interviews like Interview 1.1 and Interview 2.1 and so on. 


\emph{ Translation of Codes into themes:} After all data was provided several codes we 
have generated. All the codes were translated into several themes according the 
information. Our main in translating the coded information into themes was to obtain all 
similar information under one theme. This will also help us in analyzing the information 
which we collected.
\emph{Mapping of Themes:} Mapping of themes is the process which acted as a check 
point for the standard of information which we have collected. This assisted to assessing 
if the amount of information is sufficient for our research and also checks on if we have 
missed out on any aspect during our process. All the themed information is mapped with 
the relevant codes during this process.
\emph{Assess the trustworthiness of our synthesis:}This process is to assess that if we 
had achieved our anticipated results and are the results obtained after the thematic 
analysis are in sync in what we actually desired. This also helped us in gaining confidence 
when we know that our analysis came out well and this analysis is going to contribute us 
a lot in advanced stages of our research.

\subsection{Threats to validity}

\emph{Internal validity:} Before designing the questionnaire, the objectives of conducting 
an interview have been clearly defined. The literature review was conducted prior to the 
interviews as input to the interview design. Interviews were recorded reducing the risk of 
misinterpretation or missing important information while taking notes. As the interview 
was semi-structured the risk of interviewees misunderstanding questions was reduced 
given the dialog that took place between interviewers and the interviewee.

\emph{External validity:}  A different set of researchers may have different experiences 
and views of how to conduct surveys. We reduced the threat by conducting an extensive 
review of the literature overall including more than 70 references. We assured that we 
included researchers of different experience levels included novice researchers (PhD 
students who had 3-4 years of experience); experienced researchers (8-10 years of 
experience) and very experienced researchers (who had 30 years of experience). 

\emph{Construct validity:} While coding interviews data, chances are that we might have 
wrongly interpreted and coded the results. To mitigate this threat, the data after coding 
was crosschecked with the actual descriptions from interviews. Furthermore, the coding 
and structuring into higher level categories were reviewed by multiple authors. This 
increased the trust in using and interpreting the constructs described in the interviews 
correctly.

\emph{Conclusion validity:} Wrong conclusions may be drawn given the data. To reduce 
this threat multiple researchers were involved in the interpretation of the data. To also 
increase the reliability in the data we made sure that all the information obtained during 
interviews is documented immediately: \textit{``As soon after the interview as possible, to 
ensure that reflections remain fresh, researchers should review their field notes and 
expand on their initial impressions of the interaction with more considered comments and 
perceptions \cite{halcomb_is_2006}.''}

\section{Interview results}
\label{sec:results}

\subsection{Target audience and sampling frame definition and sampling plan}

\textbf{IP01. Insufficient sampling:} All the interviewers have one thing in common, 
they strongly believe that everyone who claims to use random and stratified sampling 
have actually done convenience sampling, the reason for this being in-feasibility to get a representative sample of the population. The main reason behind this is 
researchers cannot explicitly define the target population as all relevant variables characterizing the population are high in number and possibly not obtainable. There is no hard and fast rule for determining the desired sample size of a survey. It depends on 
various factors like the type of research, the researcher, population size, and sampling method. Also the respondents selected using 
random sample lack motivation as they might not know what for the survey is being done, 
or they might misinterpret the survey. Similarily, stratified sampling is 
believed to be challenging, expensive and time consuming, as the theoretical basis for defining a proper ``strata'' from the given population is missing. Also the timing factor of when the sample is obtained plays a role as the applicability of the findings. Thus, the value of the survey diminishes over time, as survey is just a snapshot of a particular situation at a specific point in time. Multiple strategies for sampling and obtaining the responses have been presented during the interviews.

\begin{itemize}
\item \textit{IS01: Use random convenience sampling:} Random convenience sampling was described as obtaining a sampling frame from personal contacts and randomly sampling form the frame. 

\item \textit{IS02: Use convenience snowball sampling:} Due to self-selection process which is 
followed by them all of them recommended the usage of convenience snowballing. In 
convenience snowballing the population characteristics are known before-hand, 
researchers select respondents based on their choice. Questionnaire is then filled and the 
respondents are asked to forward it to their peers. This way responses of high quality responses are 
obtained. Convenience snowballing can facilitate an additional number of responses if 
extended to LinkedIn, most visited blogs and forums. Posting and re posting the survey 
link in such social networks will make it be on the top and helps to obtain diversified 
responses.

\item \textit{IS03: Strive for heterogeneous sample:} heterogeneous sample, based on existing literature and your requirements

\item \textit{IS04: Characterize sample through demographic questions:} Demographic 
question helps to easily categorize the obtain data. Proper analysis method and reporting 
helps researcher to generalize the results involving some constraints.

\item \textit{IS05: Brevity:} A questionnaire should be short and precise. It must 
have a balance between time and number of questions. Interruptions might occur while 
answering the questionnaire, researchers should expect this while designing a survey.  Survey time and questionnaire length must be specified 
beforehand. Interviews longer than 20 minutes fail to get responses. The interviewee suggested a length of 10-15 or less.  They encouraged the
inclusion of a feature where respondents can pause and continue the survey, while count-down timers should not be used.

\item \textit{IS06: Attend conferences:} Attending the 
conferences related to the survey domain can also increase response rate. 

\item \textit{IS07: Guarantee anonymity:} Anonymity must be guaranteed and preserved. 

\item \textit{IS08: Outcome accessibility:} Motivate the 
respondents by promising them to present the outcome of your research.

\item \textit{IS09: Avoid rewards:} Respondents must not be baited, instead they have to motivated 
on why they should perform the survey and the benefits they derive from the participation. Thus, it was recommended to not give rewards. If using rewards they should be given at the end of the survey study to assure only receiving committed responses.  If rewards were to be given then the handover to each respondent should take place in person, though this might reduce the members of participants due to rewards. 
\end{itemize}

\subsection{Survey instrument design and evaluation}

\textbf{IP02: Flaws in the wording of questions:} Respondents may
misunderstand the context of questions, this is the common problem to every survey and 
cannot be neglected. Questions must be formulated with great care and must be understandable. 
\begin{itemize}
\item \textit{IS10: Consider question attributes:} Direct, consistent, non-contradictory, non-overlapping, and non-repeated questions 
must be asked to obtain vital information. A survey should have both open-ended and close-ended 
questions. Close ended save time and are easy for analysis, but open-ended give deeper 
insights about the study. Open ended answers also show the respondents commitment.
\item \textit{IS11: Survey pre-test:} At first an internal 
internal evaluation of the survey with research colleagues should take place followed by piloting with practitioners. Piloting the survey with 5 to 10 people helps to 
design the survey clearly. 
\item \textit{IS13: Researcher accessibility:} Researcher must be 
approachable if there are any doubts about the questions that need to be clarified.
\end{itemize}

\textbf{IP03: Likert Scale Problems:} Improper usage of Likert scale confuses the 
respondents. 
\begin{itemize}
\item \textit{IS14: Informed scale type decision:} Researchers need to investigate potential weaknesses of using 
different scales. Odd scales provide the respondent with the ability to be neutral by choosing the middle point of the scale, while even scales force the respondent to indicate a preference. The five-point Likert scale was suggested to be used due to its common usage in the information technology domain..
\end{itemize}

\textbf{IP04: Biases due to Question-Order Effect:} This effect should be addressed in a 
survey. 
\begin{itemize}
 \item \textit{IS15: Natural actions-sequence:} Randomizing the questions will not always work in software engineering because logical 
adherence might be lost. Only if the questions (or groups of questions in a branch) are self-contained then randomization can be done. Though, one should always consider that 
respondent might lose the context of the questions when randomizing.
\end{itemize}

\textbf{IP05: Evaluation Apprehension:}  Respondents expect to be 
anonymous when answering surveys with questions focusing on their assessment or questions that are personal in nature. Respondents also check for credibility of source 
while answering these questions. 
\begin{itemize}
	\item \textit{IS16: Avoid sensitive questions:} Whenever possible these kind of questions must be generally avoided, if asked they should be placed at the end and be optional. Questions must be framed in such a way that the feeling of being assessed is masked for the respondents.
	\item \textit{IS17: Include ``I do not know''-option:} By putting options like ``I don’t know'' or ``I don not want to answer'' will 
	encourage respondents to be truthful, and also it helps to rule-out inconsistent 
	responses.
\end{itemize}

\textbf{IP06: Lack of of Domain Knowledge:} This problem cannot 
be eliminated completely, and is significant in the case of open web 
surveys where survey is being answered by many unknown individuals. 
\begin{itemize}
\item \emph{IS18: Define clear criteria for sample selection:} The target population should be clearly defined and communicated in the survey. 
\item \emph{IS19: Stress the importance of honesty:} Explicitly motivate the respondents to be truthful about their experience when answering demographic questions.
\end{itemize}

\textbf{IP07: Hypothesis Guessing: } This is not  a problem in case of explanatory 
surveys. 
\begin{itemize}
	\item \emph{IS19: Stress the importance of honesty:}  Respondents 
	should not be influenced instead they should be motivated to be truthful on their part.
\item \emph{IS20: Avoid loaded questions:} Hypothesis guessing can be eliminated by not asking loaded questions.

\end{itemize}

\textbf{IP08: Translation issues:} The correct translation is one of the major problems 
when conducting global surveys. 
\begin{itemize}
	\item \emph{IS21: Collaboration with international researchers:} It is recommended to consult senior researchers who can translate the survey into their mother tongue and are from the same domain.  
		\item \emph{IS22: Avoid Google Translate:} Google 
		translate must not be used for language translations of surveys.
\end{itemize}

\textbf{P09: Cultural Issues:} Cultural issues may appear when conducting surveys globally, in particular the context may not be understood.
\begin{itemize}
	\item \textit{IS16: Avoid sensitive questions:} In particular in an unknown context it may be unknown how sensitive questions may be perceived, thus they should be avoided. 
	\item \textit{IS11: Survey pre-test:} Surveys should be pre-tested, and it may be recommended to use use face-to-face 
	interviews to gain trust of the respondents and get better insights.
	\item \textit{IS23: Use appropriate nomenclature:} Appropriate references and terms for things (e.g. concepts) should be used.
\end{itemize} 

\textbf{P10: Reliability:} It is
important to rule out the people with no hidden agenda or else they result in invalid conclusions. 
\begin{itemize}
\item \textit{IS4: Determine commitment:} In order to ensure reliability, the researchers must check whether 
the respondents are really committed towards the survey or not. One way of doing that is to use demographic or redundant questions, or to include open questions (see IS10). 
\end{itemize}

\subsection{Data analysis and conclusions}

\textbf{IP11: Response Duplication:} Response duplication needs to be detected, and will result in wrong conclusions if remaining undetected. 
\begin{itemize}
	\item \textit{IS25: Track IP address:} It can be identified and handled by 
	crosschecking IP addresses. One-time links can be sent directly to the mails, survey tools monitor the duplication. 
	\item \textit{IS26: Session cookies:} Tracking session cookies may help in detecting duplicates as well as information 
	about how many times did the respondent paused and resumed while answering. 
\end{itemize}

\textbf{IP12: Eliminating invalid responses:} The respondents may contradict themselves, which puts the validity of the survey results into question. 
\begin{itemize}
	\item \textit{IS27: Consistency checking:} During the analysis it is recommended to conduct a cross-analysis of questions using Cronbach's Alpha.
\end{itemize}

\subsection{Reporting}

\textbf{IP12: Incomplete reporting:} Incomplete reporting will result in the inability to assess and thus trust the outcomes of the survey. Two reporting items were emphasized:
\begin{itemize}
	\item \textit{IS28: Report inconsistencies:} Inconsistencies and invalid responses should not just be discarded, they have to be reported. 
	\item  \textit{IS29: Report biases:}  The researcher needs to identify relevant biases 
	and report them correctly in the study. 
\end{itemize}

\section{Discussion}
\label{sec:discussion}

\subsection{Comparison with related work}

\begin{table}[htbp]
	\centering
	\caption{Comparison of findings between literature and interviews}
	‚\includegraphics[scale=0.9]{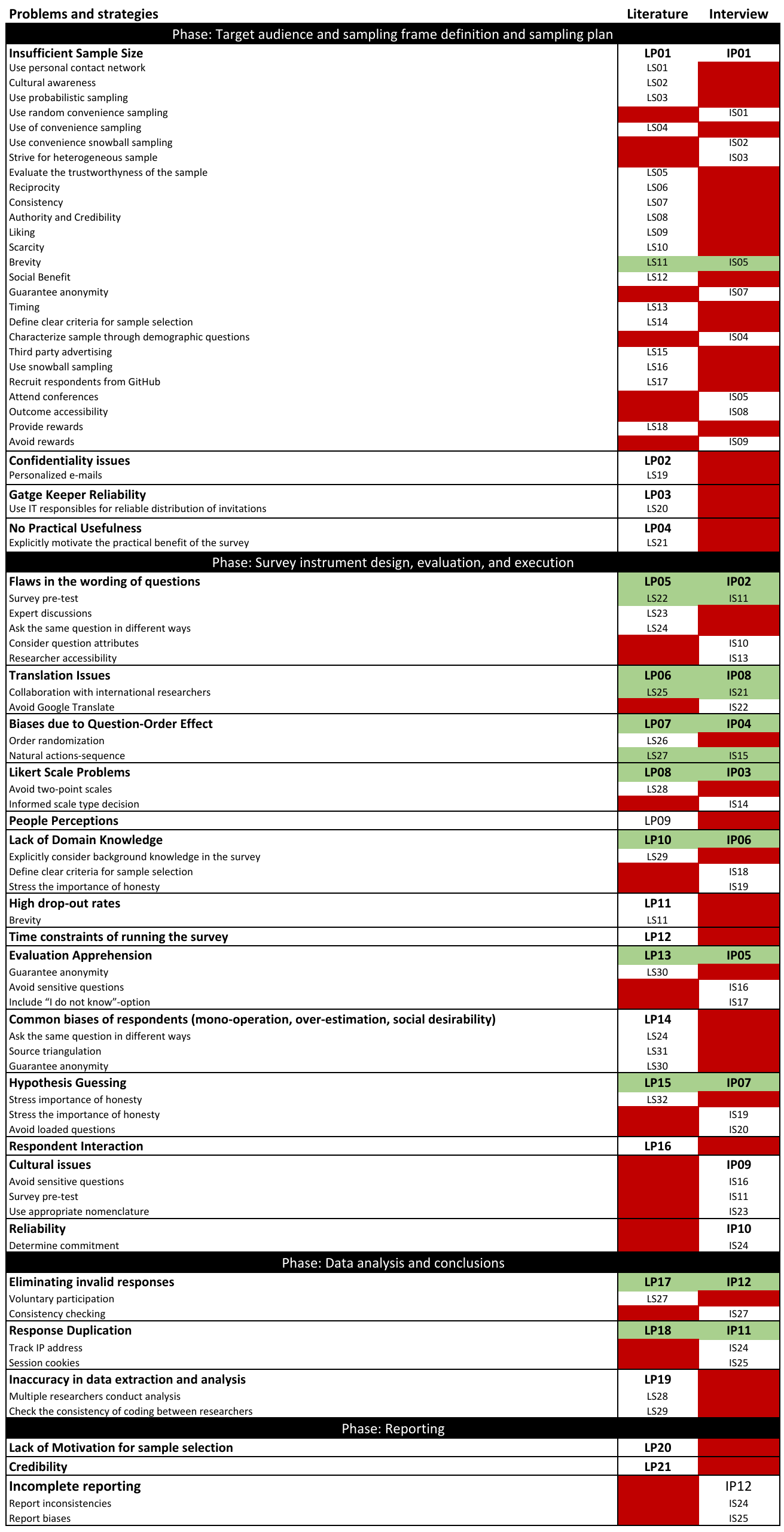}
	\label{tab:comparisonwithRW}
\end{table}

Table \ref{tab:comparisonwithRW} presents a comparison between the related work and the findings from the interviews. The problems and strategies are grouped by the phases of survey research (see Section \ref{sec:background}). Whether a problem or strategy has been identified in either the literature or interview is indicated by stating the identifiers (LP** for findings from the literature and IP** for findings from the interviews). Problems and strategies not identified by either literature or interviews are marked as ``red''; those identified by both are marked as ``green''. The table shows that literature and interviews complement each other, as each has perspective (literature or interviews) clearly shows gaps. The table may be used as a consolidated view for strategies that researchers may employ to address the problems they face during the survey research process. However, it should be noted that (a) the strategies are not validated and their effect on the quality of surveys (e.g. insufficient sample sizes) is not quantifiable. Additionally, some strategies presented in the results (Section \ref{sec:results}) and Table \ref{tab:comparisonwithRW} are conflicting, and thus designers of surveys need to make trade-off decisions when planning their research (see Section \ref{sec:tradeoffs}). Researchers conducting interviews as well as interviewees discussed incomplete reporting and the lack of motivating the sample selection. Complementary to these findings, the contents to be reported in a survey as presented by \cite{stavru2014critical} should be highlighted, which we summarized in Section \ref{sec:surveyreporting}.

\subsection{Conflicting recommendations and trade-offs}
\label{sec:tradeoffs}

Examples of conflicting strategies and the needs for trade-offs have to be highlighted considering the findings of the study. 

To address the problem of small sample sizes it was recommended to have shorter surveys (Brevity, LS11), and as questionnaire attributes the interviewees recommended non-overlapping and non-repeated questions (IS10). However, using open questions helping to determine commitment and gathering qualitative information (IS4) will make the survey longer. In addition, asking questions to check the consistency of answers (IS27, LS24) leads to a longer survey. Hence, a trade-off between the survey length reducing the number of answers and the ability to check the consistency of the survey, and gathering qualitative information needs to be made. Also, the amount of demographic information to characterize the sample (IS04) is limited when aiming for a short survey. 

Another decision concerns the type of sampling, namely probabilistic sampling (LS03) and the use of convenience sampling (LS04). As pointed out in the interviews, it is often challenging to sufficiently describe the characteristics of the population. The type of survey (exploratory versus explanatory) also influences the decision, and the degree of the ambition to generalize the survey to a population. Thus, the motivation of the sampling strategy and a clear definition of the sampling frame are essential \cite{stavru2014critical}. During the interviews hybrid strategies were identified, namely using random convenience sampling, where the list of respondents comprises of the contact networks and accessible practitioners to the researchers. From this list a random sample is then selected to partially reduce biases. 

Finally, rewards have been discussed as a strategy to increase the number of respondents. In the literature rewards were recommended as a strategy, while the risk of rewards has been pointed out (i.e. answering surveys multiple times for the sake of rewards). In the interviews it was recommended not to give rewards if mitigation strategies for addressing the risk are not addressed (e.g. receiving the rewards in person).

\section{Conclusions}
 \label{sec:conclusion}

In this study we identified problems and related strategies to overcome the problems with the aim of supporting researchers conducting software engineering surveys. The focus was on questionnaire-based research. 

We collected data from multiple sources, namely existing guidelines for survey research, primary studies conducting surveys and reporting on the problems and strategies of how to address them, as well as expert researchers. Nine expert researchers were interviewed. 

In total we identified 24 problems and 65 strategies. The problems and strategies are grouped based on the phases of the survey research process.
\begin{itemize}
	\item  \textit{Target audience and sampling frame definition and sampling plan:} It was evident that the problem of insufficient sample sizes was the most discussed problem with the highest number of strategies associated with it (26 strategies). Example strategies are brevity (limiting the length of the survey), highlighting the social benefit, using third party advertising, and the use of the personal network to recruit responses. Different sampling strategies have been discussed (e.g. random and convenience sampling). In addition more specific problems leading to losses of in responses were highlighted, such as confidentiality issues, gate-keeper reliability, and the lack of explicit motivations of the practical usefulness of the survey results. 
	\item \textit{Survey instrument design, evaluation, and execution:} The main problem observed was poor wording of questions, as well as different issues related to biases (such as question-order effect, evaluation apprehension, and mono-operation, ober-estimation, and social desirability biases). The strategies were mainly concerned with recommendations for the attributes of questions and what type of questions to avoid (e.g. loaded and sensitive questions), as well as the need for pre-testing the surveys. It was also highlighted that expert discussions are helpful in improving the survey instrument. 
	\item \textit{Data analysis and conclusions:} For data analysis the main problems were the elimination of invalid and duplicate responses as well as inaccuracy of data extraction and analysis. Technical solutions were suggested for the detection of detecting duplications. Invalid responses are avoided through consistency checking and voluntary participation. Finally, the importance of involving multiple researchers in the data analysis has been highlighted. 
	\item  \textit{Reporting:} Missing information was highlighted as problematic, including the lack of motivation for the selection of samples. It was also highlighted to report inconsistencies and biases that may have occurred in the survey. 
\end{itemize}

A high number of problems as well as strategies has been elicited. In future work a consensus building activity is needed where the community discusses which strategies are most important and suitable for software engineering research. In addition, in combination with existing guidelines the information provided in this paper may serve for the design of checklists to support the planning, conduct, and assessment of surveys.

\bibliographystyle{abbrv}

\appendix

\section{Interview guide}
\label{a:Interviewguide}

\subsection{Researcher perspective}

\begin{enumerate}
	\item You have been doing research, publishing articles since long time, by looking
	at your publications it is visible that you have conducted multiple surveys.
	Can you explain in which context you think choosing survey as a research
	method is more beneficial than action research, case studies and experiments?
	
	\item Surveys are used in social sciences and other disciplines including software
	engineering, do you think there are some special instructions to be followed
	while conducting surveys. How is it different in software engineering? What
	factors one shall consider while designing surveys in software engineering
	research?
	
	\item When designing a survey what type of questions do you prefer asking, 
	(open-ended
	or close-ended) and why? (Is the evaluation method your primary motivating factor for 
	choosing it? Are evaluation methods one of the reasons
	for choosing the type of questions? what are the other factors that enable
	you to include both type of questions in your Survey)
	
	\item In a survey one question may have provided context to the next one which
	may drive respondents to specific answers, randomization of questions to
	some extent may reduce this question-order effect. Can you suggest some
	other techniques to deal with this question order effect?
	
	\item How do you make sure that respondents understand the right context of
	your question, what measures do you adapt for making the questionnaire
	understandable?
	
	\item Our literature analysis showed that 31.4\% of primary studies used Stratified 
	sampling
	technique, while only 15.7\% of studies reported the usage of Snowball
	Sampling by researchers. (Literature describes that Snowball sampling leads
	to a better sample selection, where researcher has the freedom of choosing
	sample that suits to his/her requirements). Have you faced any situation
	where other sampling techniques were chosen over snowballing, what factors
	did you consider while making the selection?
	
	\item  Low response rates are common problem for any survey, how can the response
	rate be improved?
	
	\item  When a survey is posted, there are few respondents without a proper domain
	knowledge answering it. They might misinterpret data giving incorrect
	answers and this affects the overall analysis. In yours research how are such
	responses identified and ruled-out?
	
	\item  Our analysis showed that hypothesis guessing is an issue that can only be
	reduced to some extent rather than avoiding it completely. Explain how
	this problem is addressed in your work.
	
	\item  What measures do you take to avoid the duplication of responses in your
	surveys?
	
	\item  How do you overcome each of these common problems like bias, generalizability
	and reliability in the following cases?
	(a) Case A: Respondent’s answering the survey just for the sake of rewards.
	(b) Case B: Respondents answering surveys posted on social networks like
	LinkedIn and Facebook.
	
	\item  How do you mitigate the issue of inconsistency in responses? (Case-When a
	respondent is asked about his familiarity with non-functional requirements
	he chooses a “Yes” option. When asked to elaborate his opinion he just
	writes ``No idea'', here comes the problem of inconsistency)
	
	\item  Assume you have conducted a global survey in Sweden, Norway, China and Italy
	collecting information from diverse respondents. How do you address the
	following issues in your research?
	(a) Issue: Questionnaire gets translated into Chinese making it understandable
	to respondents over there, due to poor translation there
	might be an issue of data losses. How do you handle this language
	issue? (b) Issue: There might be some cultural issues where people of one country
	are more comfortable in answering an online questionnaire, while
	people of another country are more responsive to face-to- face interviews.
	In your opinion how can this kind cultural issue be mitigated?
	
	\item In the review of the literature we found out that researchers used Likert scale as a question type. Even though it’s a commonly used
	we could obtain few problems like central tendency bias when a 4-point or
	7-point Likert scale is used, respondent fatigue and interpretation problems
	have been identified when 9 or 10 point scales have been used. How do you
	address these kind of issues in your research?
	
	\item How do you decide upon a particular sample size for your survey?
	
	\item What motivates you to select a specific analysis technique for your research?
\end{enumerate}

\subsection{Respondent's perspective}

\begin{enumerate}
	\item You must have answered many surveys till now, in your opinion how can a
	survey grab the attention of the respondent?
	
	\item Does the questionnaire length affect your way of answering a survey?
	
	\item  What do you prefer mostly to answer, open-ended or close-ended questions?
	
	\item Does time limitation affect your way of answering the Survey?
	
	\item Would you be willing to disclose details about your research study when
	answering a survey? (Confidentiality issues, Intellectual theft)
\end{enumerate}

\subsection{Concluding questions about survey guidelines}
\begin{itemize}
	\item Do you think there is a need for the checklist of all the problems faced by
	the Software Engineering Researchers while conducting surveys?
	
	\item On the scale of 1 to 5, please rate the need for having such kind of checklist.
\end{itemize}



\end{document}